\documentclass[aps,prb,twocolumn,showpacs,amsmath]{revtex4}
\usepackage{graphicx}
\usepackage{bm}

\newcommand{\mkright}{draft (\today)} 

\begin{document}

\title{Inductances and attenuation constant for a thin-film superconducting coplanar waveguide resonator}

\author{John R.\ Clem}
\affiliation{%
   Ames Laboratory and Department of Physics and Astronomy, \\
   Iowa State University, Ames, Iowa, 50011--3160}

\date{\today}

\begin{abstract} 

The geometric, kinetic, and total inductances and the attenuation constant are theoretically analyzed for a thin-film superconducting coplanar waveguide (CPW) resonator consisting of a current-carrying central conductor, adjacent slots, and ground planes that return the current.  The analysis focuses on films of thickness $d$ obeying $d < 2\lambda$ ($\lambda$ is the London penetration depth), for which the material properties are characterized by the two-dimensional screening length $\Lambda = 2 \lambda^2/d$. Introducing a cut-off procedure that guarantees that the magnitudes of the currents in the central conductor and the ground planes are equal, new and simpler results are obtained for the kinetic inductance and the attenuation constant for small $\Lambda$.  
Exact results for arbitrary $\Lambda$ are presented for the geometric, kinetic, and total inductances in the limit of tiny slot widths, and approximate results are presented for arbitrary slot widths.  

\end{abstract}


\maketitle

\section{Introduction}
Superconducting coplanar waveguide (CPW) resonators have found increasing use in microwave integrated circuits.\cite{Shen94,Gupta96,Simons01,Zmuidzinas12}  Their applications include microwave filters in mobile and satellite communication,\cite{IEEE96,Booth99,Yoshida99a,Yoshida99b,Kanaya01} superconducting qubits,\cite{Wallraff04,Blais04,Abdumalikov08,Lindstrom07,Bourassa09} circuit quantum electrodynamics,\cite{Frunzio05,Goeppl08} radiation 
detectors,\cite{Mazin02,Day03,Zmuidzinas04,Mazin08,Vardulakis08,Cecil12} and parametric amplifiers.\cite{Tholen07}  Important characteristics of a CPW resonators are its inductances (geometric, kinetic, and total) and attenuation constant.  
Although well-known expressions for these characteristics are in common use,\cite{Shen94,Gupta96,Simons01} it is the purpose of this paper to review the assumptions underlying their derivations and to derive new expressions when these assumptions cannot be justified.

Considered here is a CPW resonator fabricated from a thin superconducting film of thickness $d$ centered on the $xz$ plane.  The central conductor ($|x| < a$, $|y| < d/2$) is of width $2a$ centered on the z axis, and the superconducting ground plane occupies the regions $|x| > b$, $|y| < d/2$.  Slots lie on either side of the central conductor in the gaps at $a < |x| < b$. 

Section \ref{ComplexFields} reviews the derivation of well-known results for the geometric, kinetic, and total inductances at low frequencies when the current is essentially all supercurrent. 
The current density in the central conductor and ground planes is assumed to be far below the depairing current density, so that the current dependence\cite{Clem12} of $\lambda$ is negligible. 
When all dimensions, including the thickness $d$, are much larger than the London penetration depth $\lambda$,  there is no penetration of the self-field into the conductors except within a distance $\lambda$ from the surface.  However, when $d < 2\lambda$,  the screening of the self-field is no longer governed by $\lambda$ but by the Pearl length\cite{Pearl64} $\Lambda = 2 \lambda^2/d$.  In this case, perpendicular magnetic fields can penetrate into the film to a distance of the order of $\Lambda$ from the film edges.  
When these depths of penetration ($\lambda$ or $\Lambda$) are much smaller than $a$, the method of complex magnetic fields\cite{Mawatari01} can be used to approximate the magnetic-field and supercurrent distributions.  I use this approach to recover the well-known result for the geometric inductance and to derive a new, simpler expression for the kinetic inductance. 
Needlessly complicated expressions for the kinetic inductance were obtained in earlier derivations assuming cutoff procedures that implicitly violated the requirement that the magnitudes of the current in the central conductor and the return current in the ground planes are the same.  

Section \ref{Pearl} contains exact solutions for the magnetic-field and supercurrent distributions and the geometric, kinetic, and total inductances  in an idealized CPW for which $b \to a$ and the screening is characterized by the 2D screening length $\Lambda$.    In Sec.\ \ref{twoWiderSlots}, I consider the realistic case of $b > a$, introduce a form for the current distribution in the central conductor that approximates its behavior as $\Lambda/a$ varies from zero to $\infty$, and obtain approximate expressions for the geometric, kinetic, and total inductances.  

In Sec.\ \ref{Losses}, I account for normal-fluid losses to derive a new, simpler expression for the CPW's attenuation constant.  Section \ref{Discussion} contains a brief summary and discussion of the results.

\section{Inductances for a thin-film CPW assuming no field penetration\label{ComplexFields}}

\begin{figure}
\includegraphics[width=8cm]{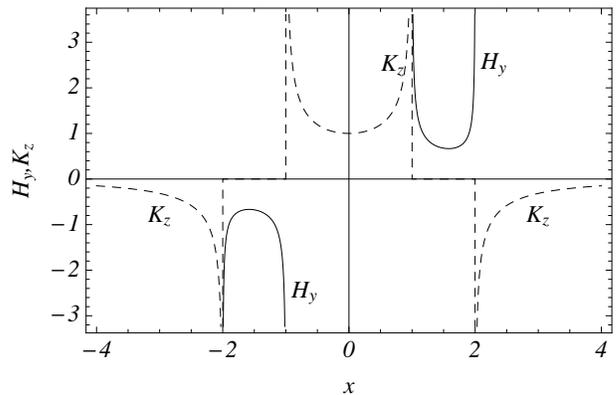}
\caption{%
$H_y(x,0)$ (solid) and $K_z(x)$ (dashed) vs $x$ for $a = 1$, $b= 2$, and $A = 1$.  }
\label{HyKzplot}
\end{figure}

\begin{figure}
\includegraphics[width=7cm]{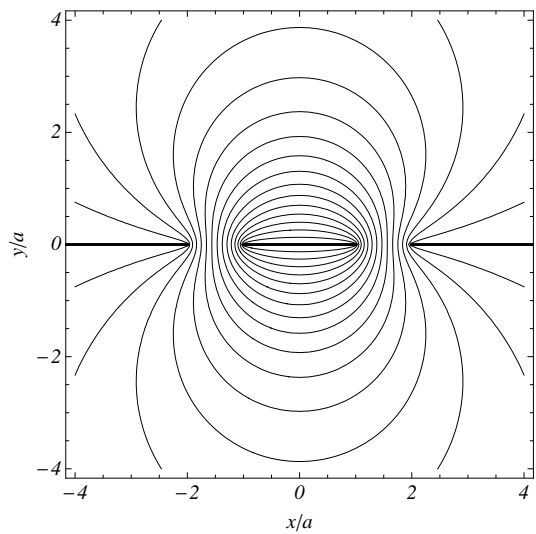}
\caption{%
Contour plot of the real part of ${\cal G}(x+iy)$ [Eq.\ (\ref{CurlyG})] as a function of $x$ and $y$, where $b = 2a$ when $\lambda^2/d  \ll a$.  The central conductor ($|x|<a$) carries current $I_z$ and the ground planes ($|x|>b$) carry the return current equally.  Contours correspond to magnetic field lines, which flow in a counterclockwise sense around the central conductor.  }
\label{G1plot}
\end{figure}

In all applications of CPWs the central conductor carries an alternating current $I_z e^{i\omega t}$.  Except at very high frequencies or at temperatures close to the transition temperature $T_c$, the current is essentially all supercurrent, and the normal-fluid contribution due to the flow of thermally excited quasiparticles is negligibly small (see Sec.\ \ref{Losses}).  To avoid including the factor $e^{i\omega t}$ for all quantities proportional to the dynamic current, let us consider the behavior at a time $t$ for which $e^{i\omega t} = 1.$ 
The central conductor ($|x| < a$, $|y| < d/2$) carries a current $I_z$ in the $z$ direction, and the semi-infinite ground planes ($|x| > b$) carry the return current in equal amounts.  When the film thickness obeys $d \ll a$ and $\lambda^2/d \ll a$, magnetic flux per unit length $\Phi'$ passes up through the right-hand slot ($a < x < b$) and down through the left-hand slot ($-b < x < -a$). We assume that $b-a \gg \lambda^2/d$, so that we can neglect the magnetic flux that penetrates into the film edges adjacent to the slots.   See Figs.\ \ref{HyKzplot} and \ref{G1plot}.

The complex magnetic field\cite{Mawatari01} ${\cal H}(\zeta) = H_y(x,y)+iH_x(x,y)$ describing this situation is, with $\zeta = x + i y$, 
\begin{equation}
{\cal H}(\zeta) =  \frac{i A}
{(\zeta^2-a^2)^{1/2}(\zeta^2-b^2)^{1/2}},
\end{equation}
where $A$ is a constant determined below.
The magnetic flux density in the plane $y = 0$ in the slots ($a < |x| < b$) is $B_y(x,0) = \mu_0 H_y(x,0)$, where 
\begin{equation}
H_y(x,0) = \frac{A x}{|x|\sqrt{(x^2-a^2)(b^2-x^2)}},
\label{Bslot}
\end{equation}
and the average sheet-current density in the central conductor and the ground plane, averaged over their thickness,  is given by $K_z(x) = -2H_x(x,\epsilon)$,
where
\begin{eqnarray}
K_z(x)& =& \frac{2 A}{\sqrt{(a^2-x^2)(b^2-x^2)}}, \; |x|< a,
\label{Kzinductance1}\\
& =& -\frac{2 A}{\sqrt{(x^2-a^2)(x^2-b^2)}}, \; |x|>b.
\label{Kzinductance2}
\end{eqnarray}
[Figure \ref{HyKzplot} shows plots of $H_y(x,0)$ and $K_z(x)$ calculated from Eqs.\ (\ref{Bslot})-(\ref{Kzinductance2}) for $a = 1$, $b = 2$ and $A = 1$.]
For the general case, the constant $A$ can be determined from the requirement that the current carried by the central conductor is $I_z$, which yields the relation
\begin{equation}
I_z = (4 A /b){\bm K}(k),
\label{Iz&A}
\end{equation}
where ${\bm K}(k)$ is the  complete elliptic integral of the first kind of modulus $k = a/b$.  
The magnetic flux per unit length $\Phi'$ can be obtained by integrating Eq. (\ref{Bslot}) over the slot, and the result is
\begin{equation}
\Phi' = \frac{\mu_0 A}{b}{\bm K}(k')= \frac{\mu_0 {\bm K}(k')}{4 {\bm K}(k)}I_z,
\label{Phi'}
\end{equation}
where $k' =\sqrt{1-k^2}$ is the complementary modulus.

The complex potential from which ${\cal H}(\zeta)$ can be derived via ${\cal H}(\zeta)=d{\cal G}(\zeta)/d\zeta$ is
\begin{equation}
{\cal G}(\zeta)=(iA/b)F(\arcsin(\zeta/a),a/b),
\label{CurlyG}
\end{equation}
where $F(\phi,k)$ is the elliptic integral of the first kind of amplitude $\phi$ and modulus $k$.
The contours in a contour plot of the real part of ${\cal G}$ correspond to magnetic field lines, as shown in Fig.\ \ref{G1plot}. These contours also correspond to contours of the vector potential $\bm A = \hat z A_z(x,y)$ ($\bm B = \nabla \times \bm A$), because $A_z(x,y) = -\mu_0 {\rm Re} {\cal G}(x+iy)$.

\begin{figure}
\includegraphics[width=8cm]{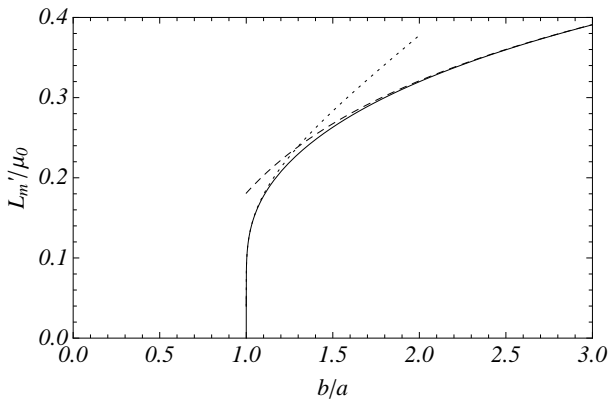}
\caption{%
$L_m'/\mu_0$, the geometric inductance per unit length in units of $\mu_0$, vs $b/a$, calculated for the case that $\lambda^2/d  \ll a$ from the exact expression Eq.\ (\ref{Lp}) (solid) and the approximations  Eq.\ (\ref{Lpapproxlarge}) (dashed) and  Eq.\ (\ref{Lpapproxsmall}) (dotted). }
\label{Lmplot1}
\end{figure} 

\subsection{Geometric inductance for $\lambda^2/d \ll a$ \label{ComplexGeometric}}

From Eqs.\ (\ref{Iz&A}) and (\ref{Phi'}) we recover the well-known result for the geometric inductance per unit length\cite{Rauch93,Watanabe94}
\begin{equation}
L_m' = \frac{\Phi'}{I_z} = \frac{\mu_0 {\bm K}(k')}{4 {\bm K}(k)},
\label{Lp}
\end{equation}
which is well approximated by 
\begin{equation}
L_m' \approx \frac{\mu_0 }{2\pi}[\ln(4b/a)-(a/2b)^2]
\label{Lpapproxlarge}
\end{equation}
for large $b/a$ and by 
\begin{equation}
L_m' \approx \Big(\frac{\pi \mu_0 }{4}\Big)/\ln\Big(\frac{8a}{b-a}\Big)
\label{Lpapproxsmall}
\end{equation}
for $(b-a)\ll a$.  
Shown in Fig.\ \ref{Lmplot1} are plots of $L_m'/\mu_0$ calculated from Eqs.\ (\ref{Lp}) (solid),  (\ref{Lpapproxlarge}) (dashed), and (\ref{Lpapproxsmall}) (dotted).

\subsection{Kinetic inductance for $\lambda^2/d  \ll a$\label{ComplexKinetic}}

For the calculations of this paper  the current density is assumed to be everywhere far below the depairing critical current density, such that only the weak-field (low-current) penetration depth $\lambda$ is involved.  For this case, since the kinetic energy per unit volume of the supercurrent is\cite{Tinkham96} $\mu_0 \lambda^2 j^2/2$, where $j$ is the supercurrent density, the kinetic inductance per unit length $L_k'$ can be calculated from
\begin{equation}
\frac{1}{2}L_k'I_z^2 = \frac{1}{2}\mu_0\lambda^2\int j^2 dS,
\label{LkDef}
\end{equation} 
where the integral is over the cross section of the CPW.  

When $\lambda^2/d  \ll a$,  the supercurrent density is given to good approximation by 
\begin{equation}
j_z(x,y) = [K_z(x)/2\lambda]\cosh(y/\lambda)/\sinh(d/2\lambda)
\end{equation}
except very close to the slots.
Here $K_z(x)$ is given by Eqs.\ (\ref{Kzinductance1}) and (\ref{Kzinductance2}), and the integral of $j_z^2(x,y)$ over the film thickness is 
\begin{equation}
\int_{-d/2}^{d/2}j_z^2 (x,y)dy = K_z^2(x)q(d/\lambda)/\lambda,
\end{equation}
where
\begin{equation}
q(u)= (\sinh u + u)/8 \sinh^2(u/2).
\label{q}
\end{equation}
Note that $q(u) \approx 1/u$ when $u \ll 1$, and $q(u) \approx 1/4$ when $u \gg 1$. 

Because $K_z^2(x)$ [Eqs.\ (\ref{Kzinductance1}) and (\ref{Kzinductance2})] diverges at the edges of the slots, the integral of this quantity over $x$ can be evaluated only approximately by cutting off the integrals at $x = \pm(a-\delta_a)$ and $x = \pm(b+ \delta_b)$,  where $\delta_a \ll a$ and $\delta_b \ll b$.  The correct cutoff procedure is to choose $\delta_a$ and $\delta_b$ such  that the magnitudes of the current in the central conductor and the total return current in the ground planes are the same.  Using this criterion to evaluate the currents for $|x|<a-\delta_a$ and $|x|>b+\delta_b$ using Eqs.\ (\ref{Kzinductance1}) and (\ref{Kzinductance2}) leads to the condition that
\begin{equation}
F\Big(\arcsin\Big(\frac{a-\delta_a}{a}\Big),\frac{a}{b}\Big) =F\Big(\arcsin\Big(\frac{b}{b+\delta_b}\Big),\frac{a}{b}\Big)  
\end{equation}
or $\delta_a/a = \delta_b/b = \epsilon \ll 1$.  
Applying this cutoff procedure to evaluate the integral of  $K_z^2(x)$ over $x$ yields, to lowest order in $\epsilon$,
\begin{eqnarray}
L_k'&=&\frac{\mu_0 \lambda q(d/\lambda)}{4 a(1-k)\bm K^2(k)}\ln\Big[\frac{2(1-k)}{\epsilon (1+k)}\Big],
\label{Lkprime}
\end{eqnarray}
where $k = a/b$.  Equation (\ref{Lkprime}) also can be written as
\begin{equation}
L_k'=\frac{\mu_0 \lambda}{2a} q(d/\lambda)g_\epsilon(k,\epsilon),
\label{Lkprime2}
\end{equation}
where
\begin{eqnarray}
g_\epsilon(k,\epsilon)&=&\frac{1}{2(1-k)\bm K^2(k)}\ln\Big[\frac{2(1-k)}{\epsilon (1+k)}\Big]
\label{gepsilon}
\end{eqnarray}
is dimensionless and for typical values of $a$ and $b$ is  of order unity (to logarithmic accuracy). 

In Refs.\ \onlinecite{Watanabe94} and \onlinecite{Booth01} the two cutoff lengths  $\delta_a$ and $\delta_b$ were assumed to be the same.  This assumption violates the requirement that  the calculated central-conductor current be the same as the magnitude of the return current in the ground plane.  Both these papers\cite{Watanabe94,Booth01} derived expressions for $L_k'$ that contain two logarithmic terms arising from the cutoffs near $a$ and $b$.  Each of the two terms resembles the logarithmic term on the right-hand side of Eq.\ (\ref{Lkprime}).  Despite differences when $b > a$,  the dependence of Eq.\ (\ref{Lkprime}) upon $a$, $b$, $k = a/b$, and the cutoff parameter when $(b-a) \ll a$ is in approximate agreement  with the more complicated Eq.\ (8) in Ref.\ \onlinecite{Booth01} for the internal inductance of a superconducting coplanar waveguide.  However, what we here call the cutoff lengths $\delta_a$ and $\delta_b$  was called the stopping distance $\Delta_x$ in Ref.\ \onlinecite{Booth01}. 

 Watanabe et al.\cite{Watanabe94} did a similar calculation using $w$ as the width of the center conductor, $s$ as the slot width, and $\delta = d/4$ as the single cutoff length.  With the replacements $a = w/2$, $b = w/2 + s$, and $\delta = d/4$,  the above expression for $g_{\epsilon}$ [Eq.\ (\ref{gepsilon})] agrees with the more complicated expression for $g(s,w,d)$ given in Refs.\ \onlinecite{Watanabe94} and \onlinecite{Yoshida95}  when $s \ll w$, except    
for a typographical error there: the $k^2$ in the denominator of the prefactor in their $g(s,w,d)$ should have been $k'^2 = 1-k^2$. 

Meservey and Tedrow\cite{Meservey69} carried out a calculation of the kinetic inductance of a long, thin superconducting strip,  and to handle a similar divergence they introduced a single cutoff length $\delta $ (similar to $\delta_a$ or $\delta_b$) given by $\delta \approx \lambda/2$ when $d  > 2\lambda$ and
 $\delta \approx d/4$ when $d  < 2 \lambda$. However, it is now well known that for $d < 2\lambda$, the equations governing the current density and self-field no longer contain $\lambda$ but the two-dimensional screening length (or Pearl length\cite{Pearl64}) $\Lambda = 2\lambda^2/d$.  (See also Sec.\ \ref{Losses}.)  As a consequence, when $d < 2\lambda$, it is incorrect to assume  $\delta \approx d/4$, and  instead one should use cutoff lengths of order $\lambda^2/d$. 

As seen from Eqs.\ (\ref{Lp})-(\ref{Lpapproxsmall}) and Fig.\ \ref{Lmplot1}, for typical values of $a$ and $b$, the geometric inductance per unit length $L_m'$ is of the order of $\mu_0$.  By comparison, the kinetic inductance per unit length $L_k'$ given in Eq.\ (\ref{Lkprime2}) depends strongly upon the ratio of $d/\lambda$. When $\lambda \ll d \ll a$, which satisfies the assumptions leading to the results of Sec.\  \ref{ComplexGeometric}, $q(d/\lambda) \approx 1/4$  [Eq.\ (\ref{q})], and $L_k' \approx (\mu_0 \lambda/8a)g_\epsilon(k,\epsilon) \ll L_m'$.  
On the other hand, when $\lambda \gg d,$ $q(d/\lambda) \approx \lambda/d$, so that Eq.\ (\ref{Lkprime2}) becomes $L_k' \approx (\mu_0 \Lambda/4a)g_\epsilon(k,\epsilon)$, in basic agreement with Eq.\ (3) of Ref.\ \onlinecite{Watanabe94} when $s \ll w$, except that the cutoff length should be $\delta \approx \lambda^2/d$ instead of $d/4$, as assumed in Ref.\ \onlinecite{Watanabe94}.

\subsection{Inductances for arbitrary $\lambda^2/d$ relative to $a$\label{ArbLambda}}

In this paper we explore how to determine both the geometric inductance and the kinetic inductance in thin films ($d < 2\lambda$) for arbitrary values of $\Lambda = 2\lambda^2/d$ relative to the width $2a$ of the center conductor.  In particular, we are also interested in examining the condition $\Lambda \gg 2a$, when the kinetic inductance is expected to be much larger than the geometric inductance.  Note that the equation $L_k' \approx (\mu_0 \Lambda/4a)g_\epsilon(k,\epsilon)$ cannot be used for this case, because it is a good approximation only when both $d \ll \lambda$ and $\epsilon \approx \lambda^2/da \ll 1$.  Moreover, Eq.\ (\ref{Lp}) for $L_m'$ is no longer valid, because its derivation required the assumption that $\lambda^2/d \ll (b-a)$, so that in the plane of the film the current-generated magnetic field was confined entirely to the slots. Therefore, to deal with thin films for arbitrary values of $\Lambda = 2\lambda^2/d$, including the case for which $\Lambda > 2a$, a  new theoretical approach  must  be employed to obtain good estimates of $L_m'$ and $L_k'$.

\section{Thin-film CPW with  narrow slits\label{Pearl}}

When the thickness $d$ of a superconducting film centered on the $xz$ plane is somewhat less than the London penetration depth $\lambda$ and the film locally carries a supercurrent density ${\bm j}(x,y,z)e^{i\omega t}$, solutions of the London equation\cite{London61} and Maxwell's equations reveal that ${\bm j}(x,y,z)e^{i\omega t}$ is practically uniform across the thickness ({\it i.e.,} independent of $y$).  One then may deal only with the sheet-current density $\bm K = \bm j d$, so that the London equation becomes $\bm K = -(2/\mu_0 \Lambda)[\bm A + (\phi_0/2\pi)\nabla \gamma],$ where $\Lambda = 2\lambda^2/d$, $\bm A$ is the vector potential in the superconducting film, $\phi_0 = h/2e$ is the superconducting flux quantum, and $\gamma$ is the phase of the superconducting order parameter.\cite{Tinkham96} The  gauge-invariant quantity $\bm A_s =\bm A + (\phi_0/2\pi)\nabla \gamma$, which is proportional to the superfluid velocity, is sometimes called the gauge-invariant vector potential. In Secs.\ \ref{oneSlit} and \ref{twoSlits}, I use  this thin-film London equation and Maxwell's equations to obtain exact analytic solutions for the geometric and kinetic inductances per unit length of a CPW resonator with very small slit widths.  As in Sec.\ \ref{ComplexFields}, to avoid including the factor $e^{i\omega t}$ for all quantities proportional to the dynamic current, it is convenient to consider the behavior at a time $t$ for which $e^{i\omega t} = 1.$

\subsection{Exact solution for a film with one long slit \label{oneSlit}}

Consider a thin superconducting film in the $xz$ plane with a long and  narrow slit along the $z$ axis containing $N'$ flux quanta per unit length, such that the magnetic flux per unit length up through the film is $\Phi' = N'\phi_0$.  We choose a gauge such that the vector potential has only a component along the $z$ direction given by 
\begin{equation}
A_0(x,y) = \int_0^\infty \alpha(k)\sin kx \;e^{-k|y|} dk.
\label{A0}
\end{equation}
Similar expressions can be obtained for the corresponding magnetic flux density components $B_{0x}(x,y)= \partial A_0(x,y)/\partial y$ and $B_{0y}(x,y)= -\partial A_0(x,y)/\partial x$.
The supercurrent sheet-current density has only a component along the $z$ direction given by the thin-film London equation,
\begin{equation}
K_0(x) = -(2/\mu_0 \Lambda)[A_0(x,0)+(\phi_0/2\pi)\gamma_z],
\label{K0Eq}
\end{equation}
where $\gamma_z(x) = \partial\gamma(x,z)/\partial z=  (\pi \Phi'/\phi_0){\rm sgn} (x)$ with ${\rm sgn} (x)=1$ for $x > 0$ and -1 for $x < 0$.  According to Ampere's law, the sheet-current density is given by the discontinuity in $B_x$ across the film thickness.  Using $K_0(x) = [B_x(x,0^-)-B_x(x,0^+)]/\mu_0$ and Eq.\ (\ref{A0}), substituting into Eq.\ (\ref{K0Eq}), and taking the derivative with respect to $x$ yields the equation  
\begin{equation}
\int_0^\infty \alpha(k)k(1+k\Lambda)\cos kx\; dk=-\Phi' \delta(x),
\label{alphaInt}
\end{equation}
where $\delta(x)$ is the Dirac delta function.  Multiplying by $\cos k'x$ and integrating over all $x$ yields
\begin{equation}
\alpha(k) =-\frac{\Phi'}{\pi k(1+k\Lambda)}.
\label{alpha}
\end{equation}

The magnetic flux per unit length $\Phi'$ carried up through the film can be described briefly as follows:   $B_{0y}(x,0)$ is an even  function of $x/\Lambda$ that diverges at $x = 0$ but decreases to zero as $|x| \to \infty$.  For $y>0$ ($y<0$) and $\rho = \sqrt{x^2+y^2} \gg \Lambda$, $\bm B_0 \approx \hat \rho \Phi'/\pi \rho$ ($\bm B_0 \approx -\hat \rho \Phi'/\pi \rho$), where  $\hat \rho = (\hat x x + \hat y y )/\rho$.   In other words, for $y > 0$, screening by the superconducting film causes the magnetic flux density at large distances to look like that produced by a magnetic line charge. 

The vector potential has the following properties: Along the $x$ axis, $A_0(x,0)$ is an odd function  of $x/\Lambda$, and $A_0(0,0) = 0$.    For  $\rho = \sqrt{x^2+y^2} \gg \Lambda$, $A_0 \approx -(\Phi'/\pi) \tan^{-1}(x/|y|)$, so that $A_0(\pm \infty,0) = \mp \Phi'/2$.

 The corresponding sheet-current density is 
\begin{equation}
K_0(x) =
-\frac{2\Phi' {\rm sgn}(x)}{\pi \mu_0\Lambda }f\Big(\frac{|x|}{\Lambda}\Big),\qquad
\label{K0}
\end{equation}
 where
\begin{equation}
f(u) =\int_0^\infty\frac{\sin t\;dt}{t+u}=
\int_0^\infty\frac{e^{-u t}\;dt}{t^2+1},
\label{f}
\end{equation}
$f(0) = \pi/2$, $df(u)/du = -\infty$ at $u =0$, and $f(u) \approx 1/u$ for $u \gg 1$. 

\subsection{Exact solution for a CPW with two  narrow slits \label{twoSlits}}

\begin{figure}
\includegraphics[width=7cm]{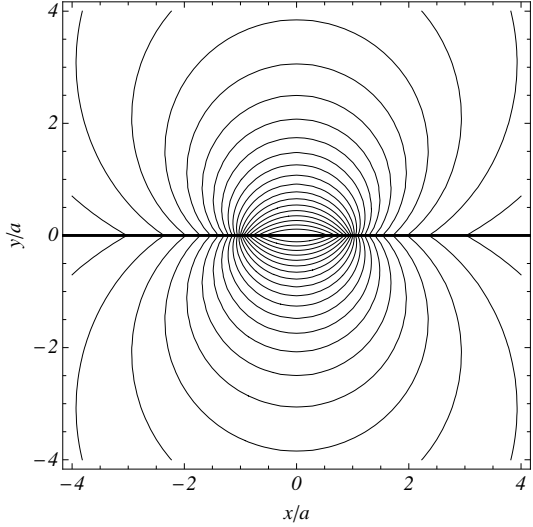}
\caption{%
Contour plot of the vector potential $A_z(x,y)$ as a function of $x$ and $y$, calculated from Eqs.\ (\ref{A0}) and (\ref{Az}) for the case of two  narrow slits at $x = \pm a$ for $\Lambda = a$. The central conductor ($|x| < a$) carries current $I_z$ and the ground plane ($|x|> a$) carries the return current.  Contours correspond to magnetic field lines, which flow in a counterclockwise sense around the central conductor.  }
\label{Az_ContourPlot}
\end{figure} 

Consider now a superconducting film in the $xz$ plane with two long and narrow slits parallel to the $z$ axis, one at $x = a$ containing $N'$ positive flux quanta per unit length, and one at $x = -a$ containing an equal number of negative flux quanta per unit length.  The resulting vector potential, magnetic flux density, and sheet-current density are obtained by a linear superposition of the reults obtained in Sec.\ \ref{oneSlit}:
\begin{eqnarray}
A_z(x,y)&=&A_0(x-a,y)-A_0(x+a,y),
\label{Az}\\
\bm B(x,y) &=&\bm B_0(x-a,y)-\bm B_0(x+a,y),
\label{B}\\
K_z(x)&=&K_0(x-a)-K_0(x+a),
\label{Kz} \\
&=&(2/\mu_0 \Lambda)[\Phi'-A_z(x,0)],\; |x|<a,
\label{Kz<}\\
&=&-(2/\mu_0 \Lambda)A_z(x,0),\; |x|>a.
\label{Kz>}
\end{eqnarray}

These fields can be described  as follows:   $B_{y}(x,0)$ is an odd  function of $x$ that diverges at $x = \pm a$ but decreases to zero as $|x| \to \infty$.   For $\rho = \sqrt{x^2+y^2} \gg \Lambda$ and $\rho \gg a$, $B_y(x,y) \approx 4\Phi'ax|y|/\pi(x^2+y^2)^2$, and $B_x(x,y) \approx \pm 2\Phi'a(x^2-y^2)/\pi(x^2+y^2)^2$, where the upper (lower) sign holds for $y>0$ ($y<0$).  In other words, for $y > 0$, screening by the superconducting film causes the magnetic flux density at large distances to look like that produced by a dipole of magnetic line charges. 

The vector potential (see Fig.\ \ref{Az_ContourPlot}) has the following properties: Along the $x$ axis, $A_z(x,0)$ is an even function  of $x$.    For  $\rho = \sqrt{x^2+y^2} \gg \Lambda$, $A_z \approx -(\Phi'/\pi) \{\tan^{-1}[(x-a)/|y|]-\tan^{-1}[(x+a)/|y|\}$, so that for $\rho = \sqrt{x^2+y^2} \gg \Lambda$ and $\rho \gg a$, $A_z(x,y) \approx 2\Phi'a|y|/\pi(x^2+y^2)$.

\begin{figure}
\includegraphics[width=8cm]{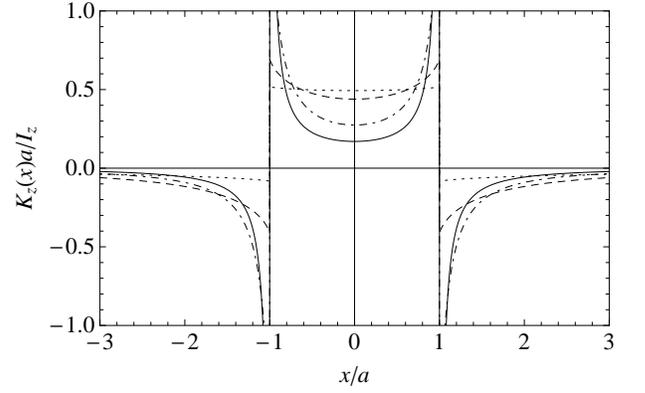}
\caption{%
Sheet-current density $K_z(x)$ from Eqs.\ (\ref{Kz})-(\ref{Kz>})], normalized to $I_z/a$ [Eq. (\ref{Iz})], vs $x/a$ for $\Lambda/a$ = 0.01 (solid), 0.1 (dot-dashed), 1 (dashed), and 10 (dotted). }
\label{KzPlot}
\end{figure}

 The corresponding sheet-current density $K_z(x)$ is an even function of $x$ with discontinuities at $x = \pm a$.   For  $\rho = \sqrt{x^2+y^2} \gg \Lambda$, $K_z(x) \approx -4\Phi'a/\pi\mu_0 x^2.$  In the limit as $\Lambda \to 0$, $K_z(x)$ reduces to the form given by Eqs.\ (\ref{Kzinductance1})-(\ref{Kzinductance2}) with $b=a$. See Fig.\ \ref{KzPlot}.

\subsubsection{Geometric inductance \label{LmtwoSlits}}

The geometric inductance per unit length $L_m'$ is defined via
\begin{equation}
\frac{1}{2}L_m'I_z^2=\int_{-\infty}^{\infty}  \int_{-\infty}^{\infty} \frac{B^2}{2\mu_0} dxdy.
\label{LmDef}
\end{equation}
The two-dimensional integral over all space can be carried out with the help of the vector identity $\nabla \cdot (\bm A \times \bm B) = \bm B \cdot \nabla \times \bm A -\bm A \cdot \nabla \times \bm B$, the divergence theorem, and  $\nabla \times \bm B = \mu_0 \bm K \delta(y).$  The surface integral at infinity vanishes, and $L_m'$ is then obtained from 
\begin{equation}
L_m'=\int_{-\infty}^{\infty}  A_z(x,0)K_z(x) dx/I_z^2,
\label{LmThin}
\end{equation}
where $A_z(x,0)$ and $K_z(x)$ are given by Eqs.\ (\ref{Az}) and (\ref{Kz}). The current $I_z$ is obtained from the integral of $K_z(x)$ over the center conductor,
\begin{equation}
I_z=\int_{-a}^{a} K_z(x) dx = \frac{4\Phi'}{\pi\mu_0}f_1(2a/\Lambda),
\label{Iz}
\end{equation}
where 
\begin{equation}
f_1(u) =\int_0^u f(u')du' =\int_0^\infty\frac{(1-e^{-ut})dt}{t(t^2+1)}.
\end{equation}

The weak dependence of $L_m'$  upon $\Lambda/a$, shown by the labeled solid curve in Fig.\ \ref{LThinPlot}, can be understood via Eq.\ (\ref{Lp}).  Although we  are dealing here with a CPW with dimensions $b/a = 1$, the penetration of magnetic flux to a distance of order $\Lambda$ to the sides of the narrow slits leads to an increase in the effective width of the slit or, equivalently, to an increase in the effective value of the ratio $b/a$. Denoting this effective value as $(b/a)_{eff} = 1/k_{eff}=(b/a)(1+c\Lambda/a)$, reveals that $c$ is a slowly varying function of $\Lambda/a$ of order unity.  For example, for the case of $b/a = 1$, substituting $(b/a)_{eff} = (1+c\Lambda/a)$ in place of $b/a$ in  Eq.\ (\ref{Lp}) and equating the result to $L_m'$ calculated from Eq.\ (\ref{LmThin}) yields $c$ vs $\Lambda/a$ as follows: $(c, \Lambda/a)$ = (0.72, 0.001), (0.68, 0.01), (0.58, 0.1), (0.37, 1), (0.20, 10), (0.14, 100), and (0.12, 1000).

\subsubsection{Kinetic inductance \label{LktwoSlits}}

The kinetic inductance per unit length $L_k'$ is defined by Eq.\ (\ref{LkDef}) and for thin films is
\begin{equation}
L_k'=\frac{\mu_0 \Lambda}{2I_z^2}\int_{-\infty}^{\infty}  K_z^2(x) dx.
\label{Lkthin}
\end{equation} 
Its behavior vs $\Lambda/a$ is shown by the labeled solid curve for $L_k'$ in Fig.\ \ref{LThinPlot}.

\subsubsection{Total inductance \label{LtwoSlits}}

As can be seen from Eqs.\ (\ref{Kz<}), (\ref{Kz>}), (\ref{Kz<}), (\ref{LmThin}), (\ref{Iz}), and (\ref{Lkthin}), the total inductance per unit length, $L' = L_m' +L_k'$ is given by
\begin{equation}
L' = \frac{\Phi'}{I_z} = \frac{\pi\mu_0}{4f_1(2a/\Lambda)}.
\label{Lthin}
\end{equation}

\begin{figure}
\includegraphics[width=8cm]{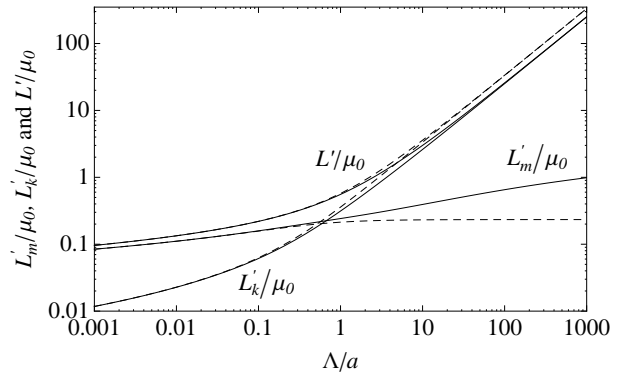}
\caption{%
The solid curves show the exact inductances per unit length (in units of $\mu_0$) calculated as in Sec.\ \ref{twoSlits}: $L'$ [total, Eq.\ (\ref{Lthin})], $L'_k$ [kinetic, Eq.\ (\ref{Lkthin})], and $L'_m$ [geometric, Eq.\ (\ref{LmThin})] for the limiting case of a thin-film CPW with two  narrow slits ($b \to a$). The dashed curves show the approximations to the corresponding inductances calculated as described in Sec.\ \ref{Approx_Narrow_Slits}. }
\label{LThinPlot}
\end{figure}

The solid curves in Fig.\ \ref{LThinPlot} exhibit how $L'_m$, $L'_k$, and $L'$ depend upon $\Lambda/a$. When $\Lambda/a \gg 1$, the inductance per unit length becomes dominated by the kinetic inductance  
contribution $\mu_0 \Lambda/4a$ arising from the nearly uniform sheet-current density $I_z/2a$ in the center conductor.

\section{Thin-film CPW with wider slots \label{twoWiderSlots}}

\subsection{General properties \label{twoWiderSlotsGeneral}}

Our goal is to calculate the inductances of a CPW for which the center conductor carries a current $I_z e^{i \omega t}$, but as in Secs.\ \ref{ComplexFields} and \ref{Pearl}, to avoid including the factor $e^{i\omega t}$ for all quantities proportional to the dynamic current, it is convenient to consider the behavior at a time $t$ for which $e^{i\omega t} = 1.$ 
  
Consider now a thin superconducting film in the $xz$ plane with two slots parallel to the $z$ axis, one at $a < x < b$ containing $N'$ positive flux quanta per unit length, and one at $-b < x < -a$ containing an equal number of negative flux quanta per unit length.  The integral of the sheet-current density $K_z(x)$ over the width $2a$ of the center conductor is $I_z$, and the integral of $K_z$ from $b$ to $\infty$ (or from $-\infty$ to $-b$) is $-I_z/2$.  Of course, $K_z$ is zero in the slots. Once $K_z(x)$ is known, the vector potential can be obtained from 
\begin{equation}
A_z(x,y) = -\frac{\mu_0}{2\pi}\int_{-\infty}^{\infty}\ln\sqrt{(x-x')^2+y^2}K_z(x') dx'.\qquad
\label{AzFromKz}
\end{equation}
Because $K_z(x)$ is also related to $A_z(x,0)$ via the thin-film London equation, we have 
\begin{eqnarray}
K_z(x)&=&
(2/\mu_0 \Lambda)[\Phi'-A_z(x,0)],\; |x|<a,
\label{Kzslots<}\\
&=&-(2/\mu_0 \Lambda)A_z(x,0),\; |x|>b,
\label{Kzslots>}
\end{eqnarray}
where $\Phi' = N'\phi_0$.
The magnetic flux-density components can be obtained from $B_x(x,y) = \partial A_z(x,y)/\partial y$ and $B_y(x,y) = -\partial A_z(x,y)/\partial x.$  
These fields have the properties that $B_{y}(x,0)$ is an odd  function of $x$ that diverges at $x = \pm a$ and $x = \pm b$ but decreases to zero as $|x| \to \infty$. $B_x(x,y)$ is an even function of $x$ but an odd function of $y$ with a discontinuity across the film at $y=0$, given by $B_x(x,0^+)-B_x(x,0^-)=-\mu_0 K_z(x)$.  For $\rho = \sqrt{x^2+y^2} \gg \Lambda$ and $\rho \gg b$, screening by the superconducting film causes the magnetic flux density at large distances to look like that produced by a dipole of magnetic line charges. 
In the limit as $\Lambda/a \to 0$, all the fields reduce to those discussed in Sec.\ \ref{ComplexFields}.

The geometric inductance per unit length $L_m'$ is defined via Eq.\ (\ref{LmDef}) and can be obtained either via  Eq.\ (\ref{LmThin}) or the double integral
\begin{equation}
L_m'=-\frac{\mu_0}{2\pi I_z^2}\int_{-\infty}^{\infty}\int_{-\infty}^{\infty}\ln|x-x'|K_z(x) K_z(x')  dxdx'.
\label{LmDouble}
\end{equation}
The kinetic inductance per unit length $L_k'$ is defined by Eq.\ (\ref{LkDef}) and can be obtained as in Eq.\ (\ref{Lkthin}). 
As can be seen from Eqs.\  (\ref{LmThin}),   (\ref{Lkthin}), (\ref{Kzslots<}), and (\ref{Kzslots>}), the total inductance per unit length, $L' = L_m' +L_k'$ is then given by
$L' = \Phi'/I_z$.

\subsection{Approximate solutions for a CPW \label{CPWApprox}}

Numerical solutions  of the thin-film London equation and Maxwell's equations would be required to calculate accurately the fields and inductances for finite gap widths $b-a$ with $b > a$.  However, it is possible to obtain reasonably good approximations to the current and field distributions for finite $\Lambda$ using the following expressions for the sheet-current density, 
\begin{eqnarray}
K_z(x)& =& \frac{2 A}{\sqrt{(a^2-p^2x^2)(b^2-p^2x^2)}}, \; |x|< a,
\label{Kzapprox1}\\
& =& 0, \;\;\;\; a< |x|< b,
\label{Kzapprox0}\\
& =& -\frac{2 A}{\sqrt{(x^2-p^2a^2)(x^2-p^2b^2)}}, \; |x|>b,
\label{Kzapprox2}
\end{eqnarray}
where the dependence of the parameter  $p$ ($0 < p < 1$) upon $\Lambda/a$ is determined as described below. This approximation automatically satisfies the requirement that $\int_{\infty}^\infty K_z(x)dx = 0$ for all $p$.  When $\Lambda/a \to 0$, we have $p \to 1$, and $K_z(x)$ is given by Eqs.\ (\ref{Kzinductance1})-(\ref{Kzinductance2}).   The constant $A$ is given for any $p$ by 
\begin{equation}
A=\frac{ b p I_z}{4F(\sin^{-1}p,a/b)},
\label{A}
\end{equation}
where $F(\phi,k)$ is the elliptic integral of the first kind of modulus $k$.  When $p \to 1$, $F(\pi/2,k)=\bm K(k),$ and we recover  Eq.\ (\ref{Iz&A}). 
When $k \to 1$, $F(\pi/2,1)=\tanh^{-1}p$.   When $\Lambda \to \infty$, $p \to 0$, $F(\sin^{-1}p,k) \to p$, and $A \to bI_z/4$.
\subsubsection{$p$ vs $\Lambda/a$ \label{pvsLambda}}

\begin{figure}
\includegraphics[width=8cm]{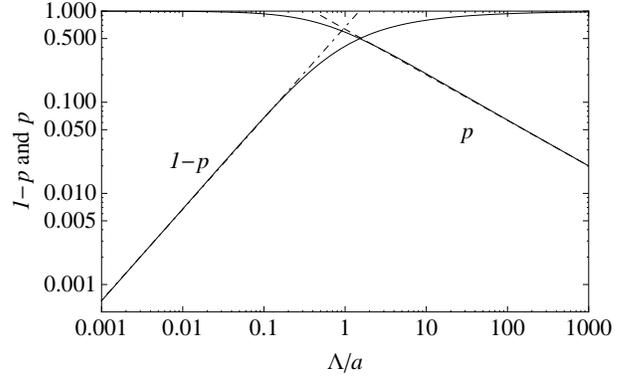}
\caption{%
The parameter $p$ and $1-p$ vs $\Lambda/a$ (solid curves) and the approximations of Eq.\ (\ref{pLargeLambda}) (dashed) and Eq.\ (\ref{pSmallLambda}) (dot-dashed).}
\label{pPlot}
\end{figure}
The dependence of $p$ upon $\Lambda/a$ shown in Fig.\ \ref{pPlot} is here obtained by equating the average over the center conductor ($-a < x < a$)  of $[K_z(x)-I_z/2a]^2$ calculated from Eq.\ (\ref{Kzapprox1})] to the average of  $[K_z(x)-I_z/2a]^2$ calculated from Eq.\  (\ref{Kz<}) for the case of $b=a$.
For large and small values of $\Lambda/a$, the numerically calculated values of $p$ and $1-p$ are well approximated by  
\begin{eqnarray}
p &\approx& 0.63/\sqrt{\Lambda/a}, \;\Lambda/a \gg 1,
\label{pLargeLambda}\\
1-p &\approx& 0.67\; \Lambda/a, \;\Lambda/a \ll 1.
\label{pSmallLambda}\\
\end{eqnarray}
It is assumed here for simplicity that the dependence of $p$ upon $\Lambda/a$ for $b > a$ remains the same as for $b = a$. 

\subsubsection{Geometric inductance \label{LmtwoSlots}}

\begin{figure}
\includegraphics[width=8cm]{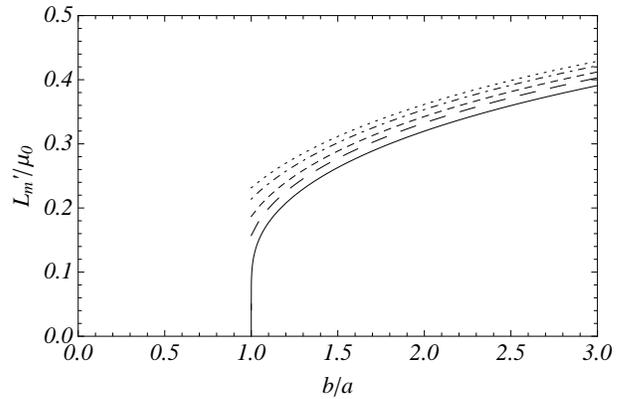}
\caption{%
$L_m'/\mu_0$, the geometric inductance per unit length in units of $\mu_0$, vs $b/a$, calculated from Eq.\ (\ref{LmDouble}) and Eqs.\ (\ref{Kzapprox1})-(\ref{A}) for $\Lambda/a = 0$ (solid, $p=1$), 0.1 (long dash, $p = 0.932$), 0.3 (dashed, $p = 0.817$), 1 (dot-dashed, $p = 0.587$),  and 10 (dotted, $p = 0.205$). }
\label{Lmplot2}
\end{figure} 

Shown in Fig.\ \ref{Lmplot2} are numerically calculated values of $L_m'$ vs $b/a$ obtained from Eqs.\ (\ref{LmDouble})-(\ref{A}) for various values of $\Lambda/a$.  For $\Lambda/a= 0$ ($p = 1$), the results reduce to those calculated in Sec.\ \ref{ComplexFields} and shown in Fig.\ \ref{Lmplot1}.  For $\Lambda/a \le 0.01$, the curves of $L_m'$ vs $b/a$ are practically indistinguishable from the solid curve, and for 
$\Lambda/a \ge 10,$ they are practically indistinguishable from the dotted curve.

Note that when $b = a$, although  $L_m' = 0$  for $\Lambda/a = 0,$ $L_m'$ is finite for all $\Lambda/a > 0,$ as shown in Fig.\ \ref{LThinPlot}.  This is due to penetration of magnetic flux through the film to a distance of order $\Lambda$ on both sides of the two slits. As discussed in Sec.\ \ref{LmtwoSlits}, $L_m'$ can be estimated from Eq.\ (\ref{Lp}) upon replacing $(b/a) = 1/k$ by $(b/a)_{eff} = 1/k_{eff}=(b/a)(1+c\Lambda/a)$, where $c$ is a slowly varying function of $\Lambda/a$ of order unity. 

\subsubsection{Kinetic inductance \label{LktwoSlots}}

The kinetic inductance is obtained by carrying out the integrals of $K_z^2(x)$ in Eq.\ (\ref{Lkthin}). Note that with Eqs.\ (\ref{Kzapprox1})-(\ref{Kzapprox2}), these integrals are convergent, so that, in contrast to the behavior discussed in Sec.\  \ref{ComplexKinetic}, no additional cutoff length needs to be introduced.
The integrals yield the following convenient analytic result for the kinetic inductance per unit length: 
\begin{equation}
L_k' = \frac{\mu_0 \Lambda}{4a}g_{kp}(k,p),
\label{LkApprox}
\end{equation}
where
\begin{equation}
g_{kp}(k,p)=\frac{(k+p^2)\tanh^{-1}p-(1+kp^2)\tanh^{-1}(kp)}{p(1-k^2)[\tanh^{-1}p]^2}
\label{gkp}
\end{equation}
and $k = a/b$. 
Note that $g_{kp}(k,1-\epsilon) \to g_\epsilon(k,\epsilon)$ in the limit as $\epsilon \to 0$. 

\begin{figure}
\includegraphics[width=7cm]{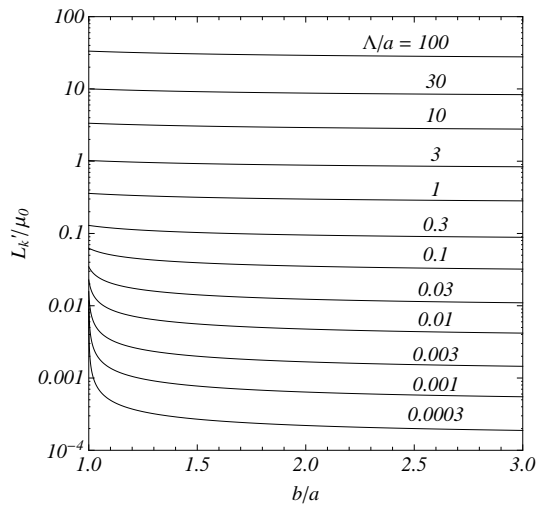}
\caption{%
$L_k'/\mu_0$, the kinetic inductance per unit length in units of $\mu_0$, vs $b/a$, calculated from Eq.\ (\ref{LkApprox}), where $p$ is calculated for each $\Lambda/a$ as in Sec.\ \ref{pvsLambda} (see Fig.\ \ref{pPlot}).}
\label{Lkplot}
\end{figure} 

Figure \ref{Lkplot} shows $L_k'$ vs $b/a$ for $\Lambda/a$ ranging from 0.0003 to 100. 

\subsubsection{Total inductance \label{LtwoSlots}}

The total inductance per unit length is given by the sum $L' = L_m' + L_k'$.   Comparison of Figs.\ \ref{Lmplot2} and \ref{Lkplot} reveals that $L_k' \ll L_m'$ when $\Lambda \ll a$, $L_k' \sim L_m'$ when $\Lambda \sim a$, and $L_k' \gg L_m'$ when $\Lambda \gg a$.  Thus, for all $b/a$, $L'$ is dominated by $L_m'$ when $\Lambda \ll a$ and by $L_k'$ when $\Lambda \gg a$, similar to the behavior of the exact solutions for $L'$ vs $\Lambda/a$ in the limit $b\to a$ shown by the solid curves in Fig.\ \ref{LThinPlot}.

\subsubsection{Comparison of approximate and exact solutions for two narrow slits \label{Approx_Narrow_Slits}}

We now examine the extent to which the inductances calculated exactly in Sec.\ \ref{twoSlits} for  two narrow slits can be  approximated by the inductances calculated using Eqs.\ (\ref{Kzapprox1})-(\ref{Kzapprox2}).  Shown as the dashed curves in Fig.\ \ref{LThinPlot} are plots of the approximate inductance contributions for  $b = a$. The kinetic inductance per unit length $L_k'$ was obtained from  Eq.\ (\ref{LkApprox}) using Eq.\ (\ref{gkp}), which in the limit $k = a/b \to 1$ reduces to
\begin{equation}
g_{kp}(1,p)=\frac{p(1+p^2)-(1-p^2)^2\tanh^{-1}p}{2p(1-p^2)[\tanh^{-1}p]^2},
\label{g1p}
\end{equation}
and $p$ was determined for each value of $\Lambda/a$ using the procedure described in Sec.\ \ref{pvsLambda}.
The geometric inductance per unit length $L_m'$ was calculated by numerically integrating Eq.\ (\ref{LmThin}) using $A_z(x)$ and $K_z(x)$ obtained from Eqs.\ (\ref{AzFromKz}) and (\ref{Kzapprox1})-(\ref{Kzapprox2}), and the total inductance per unit length was obtained from the sum $L' = L_m' + L_k'$.  

As can be seen from Fig.\ \ref{LThinPlot}, the approximate solutions quite accurately reproduce the exact results for $b =a$ for $\Lambda/a < 1$.  For large  $\Lambda/a$, however, the approximate solutions deviate from the exact results chiefly because the  magnitude of $K_z$ given by Eq.\ (\ref{Kzapprox2}) is too large for $|x|$ slightly larger than $a$, and it decreases too rapidly with $|x|$.  The exact result for $L_k'$ in the limit as $\Lambda/a \to \infty$ arises entirely from the integral of $K_z(x)^2$ over the central conductor, but the approximate calculation in the same limit gives contributions to the integral in Eq.\ (\ref{Lkthin}) from $|x|>a$ equal to 1/3 of the contribution from the central conductor.  As a result, the approximation to $L_k'$ in the limit as $\Lambda/a \to \infty$  is too large by 33\%.  The deviation of the approximation to $K_z(x)$ from the exact $K_z(x)$ for $|x| > a$ is also responsible for the fact that the approximation to $L_m'$ is too small, as shown in Fig.\ \ref{LThinPlot}.

Examination of the contributions to $L_k'$ from the integrals of $K_z^2(x)$ over $|x| < a$ and $|x| > b$ reveal that the ratio of the contribution from $|x| > b$ to the contribution from $|x| < a$ decreases as $b/a$ increases.  This suggests that the accuracy of the approximate solutions for $L'$,  $L_k'$, and $L_m'$ when  $b > a$ is better than what is shown in Fig.\ \ref{LThinPlot} for the limit $b \to a$.

\section{Accounting for normal-fluid resistive losses  \label{Losses}}

To deal with the resistive losses at frequencies of interest to electronics applications, it is useful to make use of the well-known two-fluid description,\cite{London61,Tinkham96} in which the local current density is the sum of two contributions, the superfluid current density and the normal-fluid current density $\bm j = \bm j_s + \bm j_n$.  The supercurrent density is $\bm j_s = -\bm A_s/\mu_0 \lambda^2$, where $\bm A_s$ is the gauge-invariant vector potential (see Sec.\ \ref{Pearl}) and $\lambda$ is the temperature-dependent weak-field London penetration depth.  Because the local electric field obeys $\bm E = -d\bm A_s/dt$, we have 
\begin{equation}
\bm j_s = -\frac{i}{\mu_0 \omega \lambda^2}\bm E
\end{equation}
when all fields vary as $e^{i\omega t}$. The normal-current density is
\begin{equation}
\bm j_n = \sigma_{nf} \bm E,
\end{equation}
where $\sigma_{nf} = 1/\rho_{nf}$ is the normal-fluid conductivity, {\it i.e.}, the conductivity of the thermally excited quasiparticles.  $\sigma_{nf}$ is strongly temperature-dependent, being negligibly small at temperatures $T \ll T_c$ and reducing to the normal-state conductivity $\sigma_n$ at $T = T_c$.  Note that $\bm j = \tilde \sigma \bm E$, where $\tilde \sigma = \sigma_1-i \sigma_2 = \sigma_{nf}-i/\mu_0 \omega \lambda^2$ is the complex conductivity.\cite{Tinkham96}

Consider the simple case in  which the superconductor can be approximated by a semi-infinite half-space $x > 0$ and is subjected to a parallel applied ac field in the $\hat y$ direction of amplitude $H_0$.  Making use of Ampere's law $\bm j = \nabla \times \bm H$ and Faraday's law $\nabla \times \bm E =-\mu_0 d\bm H/dt$, we find that the induced current $\bm j = \hat z j_z$ obeys 
\begin{equation}
\frac{\partial^2j_z}{\partial x^2} = \frac{j_z}{\tilde \lambda^2},
\label{jzDiffEq}
\end{equation}
whose solution is $j_z = -(H_0/\tilde \lambda)e^{-x/\tilde \lambda}e^{i\omega t}.$
Here the single material-dependent length scale in the problem is $\tilde \lambda$, the complex penetration depth\cite{Coffey91,Clem92} defined (in the absence of vortices) by
\begin{equation}
\frac{1}{\tilde \lambda^2} = \frac{1}{\lambda^2} + \frac{2i}{\delta_{nf}^2},
\end{equation}
where $\delta_{nf} = (2/\mu_0 \omega \sigma_{nf})^{1/2}$ is the normal-fluid skin depth.  Note that $\delta_{nf}$ is very large when  $T \ll T_c$ that it reduces to the normal-state skin depth $\delta_n = (2/\mu_0 \omega \sigma_n)^{1/2}$ at $T = T_c$.

Now consider the case at hand, a CPW consisting of a film of thickness $d < 2 |\tilde \lambda|$, with central conductor $|x| <a$ carrying current $I_z e^{i\omega t}$, slots at $a < |x| < b$, and ground planes carrying the return current equally  on both sides. Because the current density $j_z$ is very nearly uniform across the film thickness, it is convenient to examine the behavior of the sheet-current  density $K_z = j_z d$.  The analog of $\bm j = \tilde \sigma \bm E$ is 
\begin{equation}
K_z = -\frac{2i}{\mu_0 \omega \Lambda_\omega}E_z = Y_\omega E_z = \Big(G_{nf}-\frac{2i}{\mu_0\omega\Lambda}\Big)E_z,
\label{KzvsEz}
\end{equation}
where $\Lambda_\omega = 2\tilde \lambda^2/d$, the two-fluid version of the 2D screening length, is defined via 
\begin{equation}
\frac{1}{\Lambda_\omega} =\frac{1}{\Lambda}+\frac{i}{\Delta_{nf}},
\label{Lambdaomega}
\end{equation} 
$\Delta_{nf} = \delta_{nf}^2/d = 2/\mu_0 \omega G_{nf}$ is the normal-fluid 2D screening length, $Y_\omega$ is the complex  sheet admittance, $G_{nf} = \sigma_{nf} d = 1/R_{nf}$ is the normal-fluid sheet conductance, and $R_{nf} = \rho_{nf}/d$ is the normal-fluid sheet resistance or normal-fluid resistance per square.  Inverting  Eq.\ (\ref{KzvsEz}), we obtain 
\begin{equation}
E_z = \frac{i\mu_0 \omega \Lambda_\omega}{2}K_z = Z_\omega K_z,
\label{EzvsKz}
\end{equation}
where $Z_\omega$ is the complex sheet impedance.

In the limit as $T \to 0$, when $R_{nf} \to 0$ and $\Delta_{nf} \to \infty$, we obtain $Y_\omega \to -2i/\mu_0\omega\Lambda$, $Z_\omega \to i\mu_0 \omega \Lambda/2$, and the 2D screening length becomes the purely real quantity $\Lambda_\omega \to \Lambda = 2\lambda^2/d$, discussed in Secs.\ \ref{ComplexFields}-\ref{twoWiderSlots}.

In the limit as $T \to T_c$, when $\Lambda \to \infty$ and the normal-fluid quantities reduce to their normal-state counterparts, we see that $Y_\omega \to G_n$, $Z_\omega \to R_n = \rho_n/d$, and the 2D screening length becomes the purely imaginary quantity $\Lambda_\omega \to -i\Delta_n= -2iR_n/\mu_0\omega.$

Starting from Eq.\ (\ref{KzvsEz}) and making use of Faraday's law and the Biot-Savart law for the general case, we find that $K_z$ in the CPW obeys not a second-order differential equation such as Eq.\ (\ref{jzDiffEq}), but the following integro-differential equation (omitting the time-dependent factor $e^{i\omega t}$ in the field quantities), 
\begin{eqnarray}
\frac{\Lambda_\omega}{2}\frac{\partial K_z(x)}{\partial x} &=& H_y(x,0) \nonumber\\
&=& \frac{\rm P}{2\pi}\int_{-\infty}^\infty \frac{K_z(x')}{x-x'}dx',
\label{Kz2DEq}
\end{eqnarray}
where P denotes the principal value, and the integral is over all $x'$ except the slots ($a < |x'| < b$), where $K_z(x') = 0$.  $K_z$ is subject to the constraint that $\int_{-a}^a K_z(x) dx = I_z$.
The only material-dependent length scale in Eq.\ (\ref{Kz2DEq}) is $\Lambda_\omega$.  

Note,  however, that $K_z(x)$ and $H_y(x,0)$ vary spatially over large distances even when $\Lambda_\omega = 0$. In that case, the solutions of Eq.\ (\ref{Kz2DEq}) are given exactly by Eqs.\ (\ref{Bslot})-(\ref{Kzinductance2}), which show that  $K_z(x)$ diverges at the boundaries of the slots at $|x| = a$ and $b$.  For finite $\Lambda_\omega$,  although $\partial K_z(x)/\partial x$ and $H_y(x,0)$ still have logarithmic divergences at $|x| = a$ and $b$, $K_z(\pm a)$ and $K_z(\pm b)$ are both finite. When $|\Lambda_\omega| \ll a$, the solutions for $K_z(x)$ and $H_y(x,0)$ deviate from the $\Lambda_\omega = 0$ solutions within roughly $|\Lambda_\omega|$ of the boundaries of the slots at $|x| = a$ and $b$.

The resistive losses in the metallic components of a CPW are characterized by the attenuation constant\cite{Jackson62} $\alpha_c = (-d\bar P/dz)/2\bar P$, where $-d\bar P/dz$ is the time-averaged power dissipated in ohmic losses per unit length and $\bar P$ is the time-averaged power flow along the wave  guide. From Eq.\ (\ref{EzvsKz}) we obtain
\begin{equation}
-d\bar P/dz = -(\mu_0 \omega {\rm Im}\Lambda_\omega/4) \int_{-\infty}^\infty |K_z(x)|^2 dx,
\label{LossPerLength}
\end{equation}
and  $\bar P = I_z^2 Z_0/2$, where $Z_0$ is the characteristic impedance of the guide. 

The integral in Eq.\ (\ref{LossPerLength}) is essentially the same as that needed to calculate $L_k'$ in Eq.\ (\ref{Lkprime}).  To evaluate it approximately for $|\Lambda_\omega| \ll a$, we again cut off the integrals at $x = \pm(a-\delta_a)$ and $x = \pm(b+ \delta_b)$, where $\delta_a$ and $\delta_b$ are chosen such  that the magnitudes of the current in the central conductor and the return current in the ground plane are equal.
As in Sec.\ \ref{ComplexKinetic}, this leads to the condition that $\delta_a/a = \delta_b/b = \epsilon \ll 1$, where $\epsilon \sim  |\Lambda_\omega|/a$. The result for the attenuation constant is 
\begin{equation}
\alpha_c = \frac{R_{nf}}{8aZ_0}\Big|\frac{\Lambda_\omega}{\Delta_{nf}}\Big|^2 \frac{1}{(1-k)\bm K^2(k)}\ln\Big[\frac{2(1-k)}{\epsilon (1+k)}\Big],
\label{alphac}
\end{equation}
where $k = a/b$.   The factor $R_{nf}|\Lambda_\omega/\Delta_{nf}|^2$ reduces to zero in the limit $T \to 0$ and to $R_n$ in the limit $T \to T_c$. 

References \ \onlinecite{Booth99} and \onlinecite{Holloway95}  derived similar but more complicated expressions for $\alpha_c$ containing two logarithmic terms arising from the unequal-current assumption of a single cutoff length (called the stopping distance $\Delta$) at both $|x| = a$ and $b$. Despite numerical differences when $b > a$, the equal-current result given in Eq.\ (\ref{alphac}) is in good agreement with the results in Refs.\ \onlinecite{Booth99} and \onlinecite{Holloway95}  in the thin-film limit $d < 2|\tilde \lambda|$ when $(b-a) \ll a$.

As noted in Ref.\ \onlinecite{Gupta96}, the total microstrip losses $\alpha_T$ are the sum of both the conductor loss $\alpha_c$ and the dielectric loss $\alpha_d$ (not considered here), and both these contributions must be included to determine the quality factors of cavities built from CPWs.

\section{Discussion \label{Discussion}}

In this paper I analyzed the geometric, kinetic, and total inductances and the attenuation constant for a thin-film superconducting coplanar waveguide (CPW) resonator.  I pointed out several limitations of expressions that were derived assuming negligible penetration of perpendicular magnetic self-fields into the film.  Using a theoretical approach that guarantees equal currents in the central conductor and the ground planes, I derived new, simpler expressions for the kinetic inductance and the attenuation constant valid for small values of the ratio $\Lambda/a$ or $|\Lambda_\omega|/a$.  I derived exact results for the idealized case of a CPW in the limit as $b \to a$ for arbitrary values of $\Lambda/a$.  I then showed how to estimate the inductances using an analytic approximation that reduces to the exact results for the current density in the central conductor  in the limits $\Lambda/a \to 0$ and $\Lambda/a \to \infty$.  

For the analysis of the kinetic inductance in this paper, I assumed for simplicity that the CPW carries sufficiently low currents that suppression of the superconducting order parameter is negligible and $\lambda$ is simply the weak-field London penetration depth.  However, it is well known that application of a sufficiently large bias current can suppress the order parameter, making the effective penetration depth, the kinetic inductance, and the electrical resistance current-dependent, the changes in all these quantities initially increasing quadratically with the bias current.\cite{Anlage89,Dahm97}  Similar effects can be produced via induced currents flowing in response to an applied perpendicular field.\cite{Healey08} These properties make it possible to fabricate a variety of devices taking advantage of the associated nonlinearities.\cite{Anlage89,Lam92,Enpuku93,Cho94,Enpuku95,Tinkham96,Tholen07,Healey08,Annunziata10}

To treat current-dependent and associated nonlinear effects  was beyond the scope of this paper.  Within Ginzburg-Landau theory, the degree of order-parameter suppression scales
approximately as $(j_z/j_d)^2$, where $j_z$ is the local current density and $j_d$ is the Ginzburg-Landau depairing current density.\cite{Dahm97,Anlage89,Clem12}  As seen in Figs.\ \ref{HyKzplot} and \ref{KzPlot} and in Eqs.\ (\ref{Kzinductance1})-(\ref{Kzinductance2}), (\ref{Kz})-(\ref{Kz>}), and (\ref{Kzapprox1})-(\ref{Kzapprox2}), the current density in the CPW  generally is strongly dependent upon $x$, which would  make calculations of the current dependence of both the kinetic inductance and the dissipation very complicated, except when $\Lambda/a \gg 1,$ when the current density in the central conductor becomes practically uniform.

\section*{ACKNOWLEDGMENTS} 
I thank K. K. Berggren, D. T. Meyer, and T. P. Orlando for valuable correspondence.
This research was supported by the U.S.\ Department of
Energy, Office of Basic Energy Science, Division of Materials
Sciences and Engineering and was performed at
the Ames Laboratory, which is operated for the U.S.\ Department
of Energy by Iowa State University under Contract No.
DE-AC02-07CH11358.

\end{document}